\documentclass[aps,prl,twocolumn,floats,floatfix,english]{revtex4}
\usepackage[T1]{fontenc}
\usepackage[latin1]{inputenc}
\usepackage{amsmath}
\usepackage{graphicx}
\usepackage{setspace}
\usepackage{epsfig}
\makeatletter

\providecommand{\LyX}{L\kern-.1667em\lower.25em\hbox{Y}\kern-.125emX\@}

\usepackage{babel}
\makeatother
\begin{document}
\title{Adaptation using hybridized genetic crossover strategies}

\author{Marko Sysi-Aho}
\email[]{msysiaho@lce.hut.fi}
\author{Anirban Chakraborti}
\email[]{anirban@lce.hut.fi}
\author{Kimmo Kaski}
\email[]{Kimmo.Kaski@hut.fi}

\affiliation{Laboratory of Computational Engineering, Helsinki
University of Technology, \\
P. O. Box 9203, FIN-02015 HUT, Finland.}

\begin{abstract}

We present a simple game which mimics the complex dynamics
found in most natural and social systems. Intelligent players
modify their strategies periodically, depending on their performances.
We propose that the agents use hybridized one-point genetic crossover mechanism,
inspired by genetic evolution in biology, to modify the strategies and replace
the bad strategies. We study the performances of the agents under different 
conditions and investigate how they adapt themselves in order to survive or be 
the best, by finding new strategies using the highly effective mechanism 
we proposed. 

\end{abstract}

\maketitle
The behaviour of most of the complex systems found in natural and
social environments can be characterized by 
the competition among interacting agents for scarce resources and
their adaptation to the environment \cite{parisi,huberman,nowak,lux,arthur}.
The agents could be
diverse in form and in capability, for example, cells in an immune
system to great firms in a business centre. In these dynamically
evolving complex systems the nature of agents and their manners
differ. In order to have a deeper 
understanding of the interactions of the large number of agents, one  
should study the capabilities of the individual agents. An agent's behaviour may be 
thought of as a collection of rules governing ``responses'' to ``stimuli''.
For example, if one sees a predator, then one should run, or if the
stock indices fall then one should take immediate action, and so
on. Therefore, in order to model any complex dynamically adaptive system, a major concern is the
selection and representation of the stimuli and responses, since the behaviour
and strategies of the component agents are determined thereby. 
In a model, the rules of action are a straightforward way to describe agents' strategies.
One studies the 
behaviour of the agents by looking at the rules acting sequentially. Then
one considers ``adaptation'', which is described in biology
as a process by which an organism tries to fit itself into its environment. 
The organism's experience guides it to change its structure so that as time 
passes, the organism makes better use of the environment for its own benefit.
The timescales over which the agents adapt vary from one system to
another. For example, adaptive
changes in the immune system take hours to days, adaptive changes in a 
firm take usually months to years, and adaptive changes in the
ecosystem require years to several millennia.

In complex adaptive systems,
a major part of the environment of a particular agent includes other
adaptive agents. Thus, a considerable amount of an agent's effort goes
in adaptation to the other agents. This feature is the main source of the
interesting temporal patterns that these complex adaptive systems produce.
For example, in financial markets, human beings react with strategy and foresight by
considering outcomes that might result as a consequence of their
behaviour.This brings in a new dimension to the system, namely
rational actions, which are not innate to agents in natural
environments. To handle this new dimension, game theory is used. It helps in making
decisions when a number of rational agents are involved under conditions
of conflict and competition \cite{game}. However, game theory and
other conventional theories in
economics, study patterns in behavioural equilibrium that induce no further
interaction. These consistent patterns are quite different from the 
temporal patterns that the complex adaptive systems produce. 

In this letter, we study a simple game
which has most of the discussed features of a complex adaptive system. 
The ``mixed'' strategies
which the agents use to decide the course of action must be good,
especially when the agents have to be the best in order to survive--
similar to the idea of ``survival of the fittest'' in biology. So just as an
organism adapts itself in the natural environment, we propose that 
``intelligent'' agents in the game adapt themselves by modifying their 
strategies from time to time, depending on their current performances.
We also borrow the concept of {}``hybridization'' from biology
and use it to modify the strategies in the course of the game, in
the same way as in genetic algorithms
\cite{holland,goldberg,lawrence}. Therefore, our
game is 
a variant of the intelligent minority game introduced in \cite{Marko1},
based on the basic minority game
\cite{challet1,challet2,cavagna,riolo,lamper}. In the game we study
here, we use the mechanism of hybridized genetic crossover where
the two best strategies of an agent serve as the
{}``parents'' which it uses to create two new {}``children''
using one-point genetic crossover
\cite{lawrence,Marko1} and then replaces two of its worst strategies with
the children. 

Our game consists of an odd number of agents $N$ who can perform
only two actions denoted here by $0$ or $1$, at a given time $t$.
For example, the two actions could be ``buying'' and ``selling'' commodities/assets.
An agent wins the game if it is one of the members of the minority
group. All the agents are assumed to have access to finite amount of {}``global''
information: a common bit-string {}``memory'' of the $m$ most recent
outcomes. With this there are  $2^m$ possible ``history''
bit-strings. Now, a {}``strategy'' consists of two possible
responses, which in the binary sense are an action $0$ or the
opposite action $1$ to
each possible history bit-strings. Thus, there are $2^{2^{m}}$
possible strategies constituting the whole {}``strategy space''.
In our study, we use the {}``reduced strategy space'' by picking
only $2^{m}$ uncorrelated strategies, i.e., strategies which have
Hamming distance $d_H=1/2$ \cite{challet3}.

At the beginning
of the game, each agent randomly picks $s$ strategies which
constitutes its pool. Each time
the game has been played, time $t$ is incremented by unity and one {}``virtual'' point is assigned to a strategy that has
predicted the correct outcome and the best strategy is one which has
the highest virtual point score. The performance of the player is measured
by the number of times the player wins, and the strategy, which the
player uses to win, gets a {}``real'' point. The number of agents
who have chosen a particular action, say $1$, is denoted by
$A_{1}(t)$ and varies with time. The total utility of the system can be defined as

\begin{equation}
U(x_t)=(1-\theta(x_M))x_t+\theta(x_M)(N-x_t),
\end{equation}

\noindent where $x_M=(N-1)/2$, $x_t \in \{0,1,2,\ldots,N\} \quad
\forall t$, and

\begin{displaymath}
\theta(x_M)=\left\{ \begin{array}{ll}
              0 & \textrm{ when $x_t \le x_M$} \\
             1 & \textrm{ when $x_t > x_M$}.
\end{array}\right.
\end{displaymath}
 
When $x_t \in \{x_M,x_M+1\}$, the total
utility of the system is maximum $U_{max}$ ($=U(x_M)=U(x_M+1)$) as the highest number of players
win. The system is more efficient when the deviations from the
maximum total utility $U_{max}$ are smaller, or in other words, the fluctuations in $A_{1}(t)$
around the mean become smaller.

The players examine their performances
after every time interval $\tau $, and we call $\tau$ the
crossover time. If a player finds that he is among
the fraction $n$ (where $0<n<1$) who are the worst performing
players, he adapts himself and modifies his strategies. The mechanism by which
the player creates new strategies is that of hybridized one-point
genetic crossover, whereby 
he selects the two best strategies ({}``parents'') from his pool
of $s$ strategies. Then using one point genetic crossover
\cite{lawrence, Marko1}, he creates two new strategies ({}``children'')
and replaces his two worst strategies with the children. It should be noted that our mechanism of evolution
of strategies is considerably different from earlier attempts \cite{challet1,li1,li2}. Here, the strategies are changed by the agents themselves
and even though the strategy space evolves continuously, its size
and dimensionality remain the same.

The time variations of the number of players $A_1(t)$ who choose
action $1$ are plotted in Figure 1. We observe 
large fluctuations around the mean for the basic minority game in
Figure 1 (a). In
Figure 1 (b), we observe the effect of hybridized genetic crossovers on the
fluctuations around the mean. Interestingly, the fluctuations
disappear totally and the system stabilizes to a state where the
total utility of the system is at maximum, since at each time step
the highest number of players win the game. As expected, the 
behaviour depends on the parameter values for the system. For
example, as we increase $m$ it is more unlikely that the system
stabilizes. Also, we have to increase $s$, the size of the pool of
strategies, in order that the system stabilizes. The dependence of
the system's stability on these parameters is being studied in
details in \cite{Marko2}. So the important fact here is that the behaviour is totally
different from the behaviour of the basic minority game, where
increasing $s$ usually leads to larger deviations \cite{challet1,challet2}.
It is also interesting to note that starting from a situation,
similar to what is shown in Figure 1 (a), simply allowing the agents
to adapt themselves by modifying their strategies using the
mechanism we have proposed, drives the system towards a 
state where the total utility is optimized.

\begin{figure}
\epsfig{file=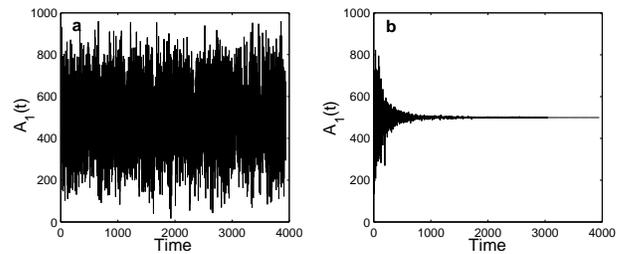,width=3.2in }
\caption{
Plot to show the time variations of the number of players $A_{1}$ who choose action 1, with the parameters $N=1001$, $m=5$, $s=10$ and $t=4000$ for (a) basic minority game and 
(b) our game, where $\tau=25$ and $n=0.6$.
}
\label{fig1}
\end{figure}

In Figure 2, we further analyze some measures related to the
simulation in Figure 1 (b). If we plot the performances of the
agents in a basic minority game, we find that the distribution of
the performances is quite symmetric around the mean and the
performances of the players do not vary remarkably during
the game \cite{challet1}. However, in our model the competition is
very stiff and there are lots of ups and downs in the performances,
and finally, when the system reaches an optimal state, the
players can be divided clearly into two groups depending on their
performances, as shown in Figure 2 (a). The performances in all
cases are scaled such that the mean performance is zero at every
time step, so that we can compare them easily. Figure 2 (b)
shows the evolution of the history. Since $m=5$, there are 
$2^m=32$ possible history bit strings denoted by a number between
$1$ and $32$. Before the
system reaches the optimal state, histories vary over the whole
range of possible outcomes as shown in Figure 2 (b). But after
reaching the stable state, the history is restricted to one value.
So, one group wins while the other loses continuously, depending on
the strategy spaces of the players. To study the differences in the
strategies of each
players' pool after the system has reached
the stable state, we have calculated the average Hamming distance
for the players' pools. Results are shown in Figure 2 (c).
If all the strategies in a pool are similar,
average Hamming distance is zero, for uncorrelated strategies, it
is $1/2$ and for totally anti-correlated strategies $1$.
Surprisingly, we find that for most of the players the average Hamming
distance calculated for the whole pool is zero, which implies that
the players have evolved their strategies and found only one
strategy for use.

\begin{figure}
\epsfig{file=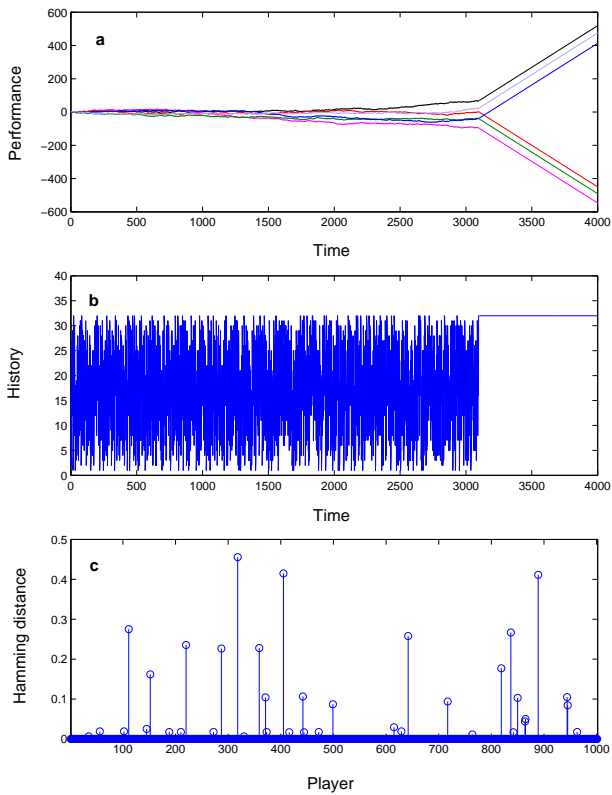,width=3.2in }
\caption{
Plots to show (a) the performances of the players in our game for the best 
player (black), the worst player (magenta) and four randomly chosen players 
(blue, green, red and violet), (b) the time variation of the history and (c) the
Hamming distances for the final pool of strategies for all the players. The parameters used for simulations are $N=1001$, $m=5$, $s=10$, $t=10000$ and $\tau=25$ and $n=0.6$.
}
\label{fig2}
\end{figure}

In order to study the efficiency of the system, 
we have introduced the study of the variation of the average total utility of the system
$U(x_t)$ with time $t$. The results are shown in Figure 3. We find that for the
basic minority game the total utility does not change much
throughout the course of the game. However, in the game we study,
we can clearly see that the total utility of the system increases
as the time passes on and eventually saturates. We can
define a characteristic time, called the ``adaptation time''
$\lambda$, during which the total utility reaches a saturation point.
As intuitively expected, the adaptation time $\lambda$ depends on the parameters of the system.
We defer the detailed studies and results for a future
communication \cite{Marko2}. It is interesting to note that this
utility measure is effective in characterizing an adaptive game.

\begin{figure}
\epsfig{file=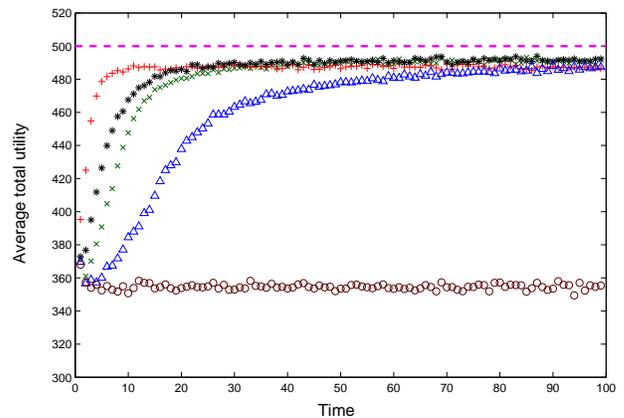,width=3.2in }
\caption{
Plot to show the variation of total utility of the system with time for the 
basic minority game for $N=1001$, $m=5$, $s=10$, $t=5000$, and our game, for 
the same parameters but different values of  $\tau$ and 
$n$. Each point represents a time average of the total utility for separate 
bins of size 50 time-steps of the game. The maximum total utility ($=(N-1)/2$
is shown as a dashed line. The data for the basic minority game is shown in 
circles. The plus signs are for $\tau=10$ and $n=0.6$; the asterisk marks are
for $\tau=50$ and $n=0.6$; the cross marks for $\tau=10$ and $n=0.2$ and 
triangles for $\tau=50$ and $n=0.2$. We have taken ensemble average over 70 
different samples, in each case.
}
\label{fig3}
\end{figure}

In order to demonstrate that the players who adapt themselves in
the course of the game, by modifying their strategies using the
hybridized one-point genetic crossover, we have tested the players
in two different situations. The first situation is where all the players play
the basic minority game but later we select the worst player and
allow it to adapt itself and thus modify its strategies. We find
that the player starts winning immediately and eventually comes out
to be a winner as shown in Figure 4 (a). Further, we choose two other worst players at two different
times and allow them to modify their strategies also. These two
players too begin to perform very well. We find that the
performances of these chosen ``intelligent'' players are much
better compared to the ``normal'' players of the basic minority
game.
The second situation consists of ten ``intelligent'' players
who are capable of modifying their strategies and the rest are
``normal'' players who simply play the basic minority game.
We find that the intelligent players perform extremely well in
comparison to the other normal players, and form a separate group, as shown in Figure 4 (b). The competition amongst
themselves is very stiff as can be seen from the inset of Figure 4 (b).

\begin{figure}
\epsfig{file=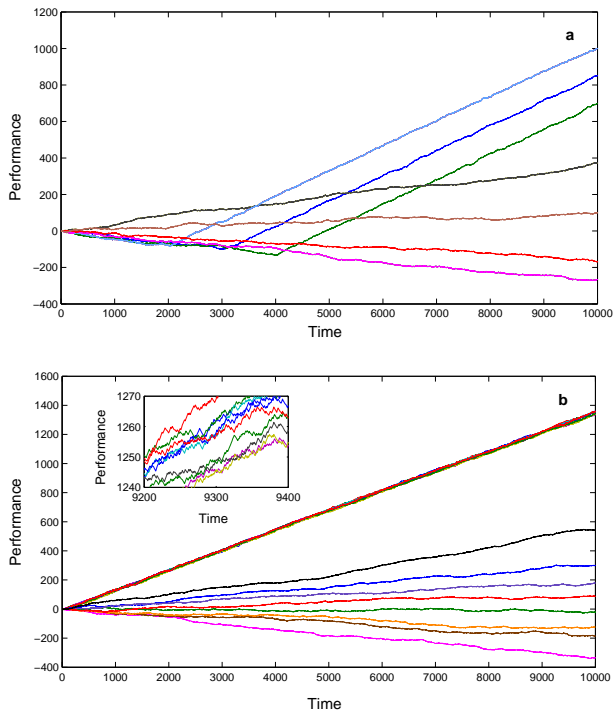,width=3.2in }
\caption{
Plots to show the performances of the players (a)  for three players who were 
the worst players in the basic minority game at different times but started winning once they started modifying 
their strategies using the hybridized genetic crossovers (light blue, navy blue and green) and the best ``normal'' player (black), the worst ``normal'' player 
(magenta) and two randomly chosen ``normal'' players (brown and red) and (b)
for the ten ``intelligent'' players who modify their strategies using the 
hybridized genetic crossovers and cluster together in the winning group and the best ``normal'' player (black), the worst ``normal'' player
(magenta) and six randomly chosen ``normal'' players (blue, violet, red, green,
orange and brown). 
The inset of (b) shows performances of ten ``intelligent'' players who modify 
their strategies using the
hybridized genetic crossovers and cluster together in the winning group in a 
magnified scale
The parameters used for simulations are $N=1001$, $m=5$, $s=10$, $t=10000$
and $\tau=10$.}
\label{fig4}
\end{figure}

These two situations and the results clearly show how effective the
adaptation of agents can be in a complex adaptive system. The
mechanism of modifying the strategies is also very successful as
it allows a player to find new strategies which maximizes
the players' individual utility. However, the total utility of the
system does not change much as the fraction of adaptive players is
very small in both cases. It would be interesting to study the
variation of the total utility of the system with the fraction of
adaptive players.

In summary, we have proposed a game where the players adapt
themselves to continuously changing environment, thus reproducing
interesting temporal patterns that are usually created by complex
adaptive systems in nature. The mechanism of adaptation we have
introduced here seems to be very effective in all the cases we have
studied, as can be seen from
the individual performances of the players or from the measure of
the total utility of the system. The performances of the players in different
conditions always seemed to be better when they adapted
themselves compared to the players who did not. We
conclude that using this mechanism one could increase remarkably the
individual utility and the total utility of the system as well, if
the fraction of adaptive players is significant.

\begin{acknowledgments}
This research was partially supported by the Academy of
Finland, Research Centre for Computational Science and Engineering,
project no. 44897 (Finnish Centre of Excellence Programme 2000-2005).
\end{acknowledgments}


\begin{thebibliography}{10}
\bibitem{parisi}G. Parisi, \textit{Physica A} \textbf{263}, 557 (1999).
\bibitem{huberman}B. A. Huberman, P. L. T. Pirolli, J. E. Pitkow and R. M. Lukose, \textit{Science}
\textbf{280}, 95 (1998).
\bibitem{nowak}M. Nowak and R. May, \textit{Nature} \textbf{359}, 826 (1992).
\bibitem{lux}T. Lux and M. Marchesi, \textit{Nature} \textbf{397}, 498 (1999).
\bibitem{arthur}W. B. Arthur, \textit{Am. Econ. Rev.} \textbf{84}, 406 (1994).
\bibitem{game}R. Myerson, \emph{Game Theory: Analysis of Conflict}
(Harvard University Press, Cambridge, Massachusetts, 1991).
\bibitem{holland}J. H. Holland, \textit{Adaptation in Natural and Artificial Systems},
University of Michigan Press, Ann Arbor (1975).
\bibitem{goldberg}D. E. Goldberg, \textit{Genetic Algorithms in Search, Optimization
and Machine Learning}, Addison-Wesley, Reading, Massachusetts (1989).
\bibitem{lawrence}D. Lawrence (Ed.), \textit{Handbook of Genetic Algorithms}, Van Nostrand
Reinhold, New York (1991).
\bibitem{Marko1}M. Sysi-Aho, A. Chakraborti and K. Kaski, \emph{preprint
available at cond-mat/0209525} (2002).
\bibitem{challet1}D. Challet and Y.-C. Zhang, \textit{Physica A} \textbf{246}, 407 (1997).
\bibitem{challet2}D. Challet, M. Marsili and R. Zecchina, \textit{Phys. Rev. Lett.} \textbf{84},
1824 (2000).
\bibitem{riolo}R. Savit, R. Manuca and R. Riolo, \textit{Phys. Rev. Lett.} \textbf{82}, 2203 (1999).
\bibitem{cavagna}A. Cavagna, J. P. Garrahan, I. Giardina and D. Sherrington, \textit{Phys. Rev. Lett.} \textbf{83}, 4429 (1999).
\bibitem{lamper}D. Lamper, S. D. Howison and N. F. Johnson, \textit{Phys. Rev. Lett.}
\textbf{88}, 17902 (2002).
\bibitem{challet3}D. Challet and Y.-C. Zhang, \textit{Physica A} \textbf{256}, 514 (1998).
\bibitem{li1}Y. Li, R. Riolo and R. Savit, \emph{Physica A} \textbf{276}, 234 (2000).
\bibitem{li2}Y. Li, R. Riolo and R. Savit, \emph{Physica A} \textbf{276}, 265 (2000).
\bibitem{Marko2}M. Sysi-Aho, A. Chakraborti and K. Kaski, in preparation (2002).
\end{thebibliography}
\end{document}